\documentclass[reprint,aps,prl,twocolumn,longbibliography,superscriptaddress]{revtex4-1}
\usepackage{graphicx}
\usepackage{amssymb}
\usepackage{amsmath}
\usepackage{bm}
\usepackage{color}
\usepackage[utf8]{inputenc}
\usepackage{chngcntr}

\begin{document}

\title{Quantum fluctuations of the center-of-mass and relative parameters of
NLS breathers}

\author{Oleksandr V. Marchukov}
\email[]{oleksandr.marchukov@tu-darmstadt.de}
\affiliation{Technische Universit\"{a}t Darmstadt, 64289 Darmstadt, Germany}
\affiliation{ Department of Physical Electronics, School of Electrical
Engineering, Faculty of Engineering, and Center for Light-Matter Interaction,
Tel Aviv University, 6997801 Tel Aviv,
Israel}
\author{Boris A. Malomed}
\affiliation{ Department of Physical Electronics, School of Electrical
Engineering, Faculty of Engineering, and Center for Light-Matter Interaction,
Tel Aviv University, 6997801 Tel Aviv,
Israel}
\affiliation{Instituto de Alta Investigaci\'{o}n, Universidad de Tarapac\'{a}, Casilla 7D, Arica, Chile}
\author{Vanja Dunjko}
\affiliation{Department of Physics, University of Massachusetts Boston,
Boston, Massachusetts 02125, USA}
\author{Joanna Ruhl}
\affiliation{Department of Physics, University of Massachusetts Boston,
Boston, Massachusetts 02125, USA}
\author{Maxim Olshanii}
\affiliation{Department of Physics, University of Massachusetts Boston,
Boston, Massachusetts 02125, USA}
\author{Randall G. Hulet}
\affiliation{Department of Physics and Astronomy, Rice University,
Houston,Texas 77005, USA}
\author{Vladimir A. Yurovsky}
\affiliation{School of Chemistry, Tel Aviv University, 6997801 Tel Aviv,
Israel}

\begin{abstract}
We study quantum fluctuations of macroscopic parameters of a nonlinear Schr%
\"{o}dinger breather --- a non-linear superposition of two solitons, which
can be created by the application of a four-fold quench of the scattering
length to the fundamental soliton in a self-attractive quasi-one-dimensional
Bose gas. The fluctuations are analyzed in the framework of the Bogoliubov
approach in the limit of a large number of atoms $N$, using two models of the
vacuum state: white noise and correlated noise. The latter model, closer to
the ab initio setting by construction, leads to a reasonable agreement,
within $20\%$ accuracy, with fluctuations of the relative velocity of
constituent solitons obtained from the exact Bethe-ansatz results [Phys.\
Rev.\ Lett.\ \textbf{119}, 220401 (2017)] in the opposite low-$N$ limit (for 
$N\leq 23$). We thus confirm for macroscopic $N$ the breather dissociation
time to be within the limits of current cold-atom experiments. Fluctuations
of soliton masses, phases, and positions are also evaluated and may have
experimental implications.
\end{abstract}

\maketitle

\paragraph{Introduction.} The nonlinear Schr\"{o}dinger equation (NLSE)\
plays a fundamental role in many areas of physics, from Langmuir waves in
plasmas~\cite{D.Fried1973} to the propagation of optical signals in
nonlinear waveguides~\cite{Hasegawa1973, Hasegawa1973b, Lai1989a, Lai1989b,
Haus1996}. A variant of the NLSE, in the form of the Gross-Pitaevskii\ equation
(GPE), provides the mean-field (MF)\ theory for rarefied Bose-Einstein
condensates (BECs). Experimentally, bright solitons predicted by the GPE
with the self-attractive nonlinearity were observed in ultracold 
${}^{7}\mathrm{Li}$ 
\cite{Strecker2002,Khaykovich2002,Strecker2003} and 
${}^{85}\mathrm{Rb}$ \cite{Cornish2006, Marchant2013} gases, in the
quasi-one-dimensional (1D) regime imposed by a cigar-shaped potential trap.
Because the GPE-based MF approximation does not include quantum
fluctuations, one needs to incorporate quantum many-body effects to achieve
a more realistic description of the system. The simplest approach is to
employ the linearization method first proposed by Bogoliubov
\cite{Bogolubov1947} in the context of superfluid quantum liquids. For more than
two decades, this method has been successfully used to describe excitations
in BECs~\cite{Griffin1995, Dalfovo1999, Leggett2001, Pethick2008}. 
Another approach deals with the Lee-Huang-Yang (LHY) corrections \cite{LHY}
to the GPE induced by quantum fluctuations around the MF states 
\cite{Petrov1,Petrov2}. The so improved GPEs produce stable 2D and 3D solitons
(including ones with embedded vorticity \cite{Raymond,Barcelona}), which have
been created in experiments with binary 
\cite{Leticia1,Leticia2,Inguscio,hetero} and single-component dipolar 
\cite{Pfau1,Pfau2} BECs.

The focusing nonlinearity in the NLSE corresponds to attractive interactions
between atoms in BEC. The NLSE in 1D without external potentials belongs to
a class of integrable systems~\cite{Zakharov1972, Ablowitz1981, Novikov1984}%
, thus maintaining infinitely many dynamical invariants and infinitely many
species of soliton solutions. The simplest one, the fundamental bright
soliton, is a localized stationary mode which can move with an arbitrary
velocity. The next-order solution, i.e., a 2-soliton, which is localized in
space and oscillates in time, being commonly called a \textit{breather}, can
be found by means of an inverse-scattering-transform \cite{Satsuma1974}.
This solution may be interpreted as a nonlinear bound state of two
fundamental solitons with a $1:3$ mass ratio and exactly zero binding 
energy~\cite{Ablowitz1981, Kivshar1989}. The 2-soliton breather can be 
created by a sudden quench of the interaction strength, namely, its
fourfold increase, starting from a
single fundamental soliton, as was predicted long ago in the analytical 
form~\cite{Satsuma1974}, and recently demonstrated experimentally in 
%V a 
BEC
%V soliton 
\cite{DiCarli2019}.

Quantum counterparts of solitons and breathers can be constructed as
superpositions of Bethe ansatz (BA) eigenstates of the corresponding quantum
problem \cite{Lai1989b} which recover MF properties in the limit of large
number of atoms ($N$). While an experimental observation of the quantum
behavior of the center-of-mass (COM) coordinate of a (macro/meso)-scopic
soliton (e.g., effects analyzed in Refs. \cite%
{drummond1994,vaughan2007,Castin2009}) remains elusive, several groups have
been making progress towards this goal \cite{Boisse2017, DiCarli2019}.
Certain quantum features of NLSE breathers, such as correlations and
squeezing \cite{schmidt2000,schmidt2000b,lee2005}, conservation laws \cite%
{Drummond2017}, development of decoherence \cite{Opanchuk2017}, and nonlocal
correlations \cite{Ng2019}, have been analyzed and discussed. The non-MF
breather-like solutions were also considered in open Bose-Hubbard,
sine-Gordon, and other models \cite{Su2015, Witthaut2011}.
Note that in the semiclassical limit the instability of quantum breathers carries over into the MF regime that was explored for NLSE in various settings in Ref.~\cite{Yanay2009}.

At the MF level, the relative velocity of the fundamental solitons whose
bound state forms the breather, is identically equal to zero, regardless of
how hot the COM state of the \textquotedblleft mother\textquotedblright\
soliton was. Thus, if the breather spontaneously splits in free space into a
pair of constituent fundamental solitons, intrinsic quantum fluctuations are
expected to be the \emph{only} cause of the fission (at the MF level,
controllable splitting of the breather can be induced by a local linear or
nonlinear repulsive potential \cite{Marchukov2019}). This expectation
suggests a way to observe the splitting as a direct manifestation of 
\emph{quantum fluctuations in a macroscopic object}, which may take place
under standard MF experimental conditions.

\begin{figure}[tbp]
\centering
\includegraphics[width=0.45\textwidth]{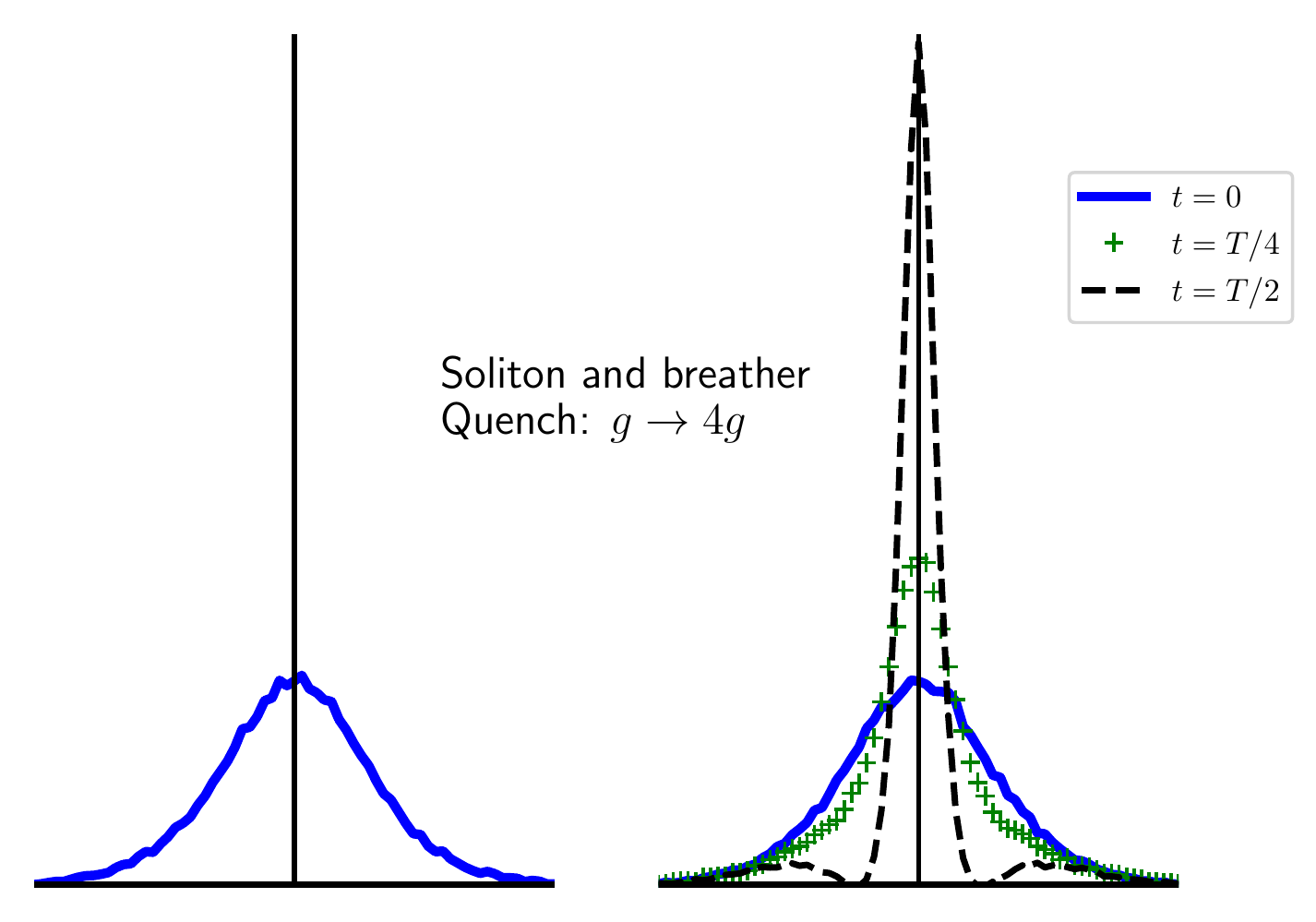}
\caption{A schematic representation of the fundamental \textquotedblleft
mother soliton", as the vacuum state including inherent correlated quantum
noise (the left panel), transformed into the breather by means of the
interaction quench (the right panel). }
\label{fig:quench}
\end{figure}

The Bogoliubov linearization method was first
applied to fundamental solitons \cite{Haus1985, Haus1990, Lai1993} in
optical fibers. Later, Yeang~\cite{Yeang1999} extended the analysis for
the COM degree of freedom of a breather. The present work focuses on quantum
fluctuations of relative parameters. 
We deal with two models for the halo of quantum fluctuations around
the MF states of the atomic BEC: conventional \textquotedblleft white
noise\textquotedblright ~ \cite{Haus1985, Haus1990,Opanchuk2017} of vacuum
fluctuations, and the most relevant scheme with correlated noise, assuming
that the breather has been created from a fluctuating fundamental soliton,
by means of the above-mentioned factor-of-four quench, as schematically
shown in Fig.~\ref{fig:quench}. For a small number of atoms, up to $N=23$,
estimates for the relative velocity variance and splitting time were
obtained in Ref. \cite{Yurovsky2017}, using the exact many-body BA solution; 
however, available techniques do not make it possible to run
experiments with such \textquotedblleft tiny solitons\textquotedblright. The present work
extends the results for the experimentally relevant large values of $N$ and
provides variances of other breather parameters, which may also be observable.

\paragraph{The system.} We consider a gas of bosons with $s$-wave
scattering length $a_{\mathrm{sc}}<0$ in an elongated trap with transverse
trapping frequency $\omega _{\perp }$ \cite{Strecker2002, Khaykovich2002,
Yurovsky2008}. The scattering length can be tuned by a magnetic field, using
the Feshbach resonance \cite{Chin2010}. Atoms with kinetic energy 
$<\hbar \omega _{\perp }$ may be considered as 1D particles with the
attractive zero-range interaction between them, of strength 
$-g=2\hbar\omega _{\perp }a_{\mathrm{sc}}$ \cite{olshanii1998}. 
The 1D gas is described by the quantum (Heisenberg's) NLSE, 
\begin{equation}
i\hbar \frac{\partial \hat{\Psi}(x,t)}{\partial t}=
-\frac{\hbar ^{2}}{2m}\frac{\partial ^{2}\hat{\Psi}(x,t)}{\partial x^{2}}
-g\hat{\Psi}^{\dagger}(x,t)\hat{\Psi}(x,t)\hat{\Psi}(x,t),  \label{qnlse}
\end{equation}%
where $m$ is the atomic mass. The creation and annihilation quantum-field
operators, $\hat{\Psi}^{\dagger }$ and $\hat{\Psi}$, obey the standard
bosonic commutation relations.

The Bogoliubov theory represents the quantum field as 
$\hat{\Psi}(x,t)=\sqrt{N}\Psi _{0}(x,t)+\delta \hat{\psi}(x,t)$, 
where the first MF term is a
solution of classical NLSE representing the condensed
part of the boson gas. Operator $\delta \hat{\psi}(x,t)$ represents quantum
fluctuations,\ also obeying the standard bosonic commutation relations. The
Bogoliubov method linearizes Eq.~\eqref{qnlse} with respect to $\delta \hat{%
\psi}$: 
\begin{equation}
i\hbar \frac{\partial \delta \hat{\psi}}{\partial t}=
-\frac{\hbar ^{2}}{2m}\frac{\partial ^{2}\delta \hat{\psi}}{\partial x^{2}}
-2gN|\Psi_{0}|^{2}\delta \hat{\psi}-gN\Psi _{0}^{2}\delta \hat{\psi}^{\dagger }.
\label{lin-qnlse}
\end{equation}%
Applying this to NLSE breathers, we use Gordon's solution of the NLSE 
\cite{Gordon1983} for two solitons with numbers of atoms 
$N_{1}$ and $N_{2}$,
which contains 8 free parameters (see 
\cite{supplement}). Four parameters represent the bosonic state as a whole: the
total number of atoms $N=N_{1}+N_{2}$, overall phase $\Theta $, COM velocity 
$V$, and COM coordinate $B$. The other four parameters are the relative velocity 
$v$ of the constituent solitons, initial distance between them, $b$, initial
phase difference, $\theta $, and mass difference, $n=N_{2}-N_{1}$. The
particular case of $n=\pm N/2$ and $v=b=\theta =0$ corresponds to the
breather solution. In the COM reference frame ($V=0$), the breather remains
localized, oscillating with period $T_{\mathrm{br}}=32\pi \hbar
^{3}/(mg^{2}N^{2})$. On the other hand, the fundamental soliton is obtained
for $n=\pm N$ and $v=b=\theta =0$.

The quantum correction to the two-soliton solution is 
\begin{equation}
\delta \hat{\psi}=\sum_{\chi }f_{\chi }(x,t)\Delta \hat{\chi}_{0}+
\hat{\psi}_{\mathrm{cont}}(x,t),  \label{quancorr}
\end{equation}%
where $\chi $ is set of the 8 parameters ($N$, $\Theta $, $V$, $B$, $n$, 
$\theta $, $v$, and $b$), and 
$f_{\chi }(x,t)=\partial (\sqrt{N}\Psi_{0})/\partial \chi $ 
are derivatives of the MF solution with respect to
them. Then, the sum in Eq. \eqref{quancorr} is an exact operator solution of
the linearized equation (\ref{lin-qnlse}). Hermitian operators 
$\Delta \hat{\chi}_{0}$, introduced in Refs. \cite{Haus1985, Haus1990}, 
are considered as
quantum fluctuations of the parameters at $t=0$, as they have the same
effect on the density as classical fluctuations of the MF parameters, see
 \cite{supplement}. The set of 8 parameters is related to breaking
of the $U(1)$ and translational symmetries of the underlying Hamiltonian,
hence they represent the Goldstone and \textquotedblleft
lost\textquotedblright\ modes, in the framework of the Bogoliubov-de Gennes
description \cite{Lewenstein1996, Castin1998, Castin2009}. The operator term 
$\hat{\psi}_{_{\mathrm{cont}}}$ in Eq. \eqref{quancorr} represents
fluctuations with a continuum spectrum (which were analyzed for the
fundamental soliton in Ref. \cite{Haus2000}). In this work, we assume
orthogonality of the continuum fluctuations $\hat{\psi}_{\mathrm{cont}}$ to
the discrete-expansion modes, leaving a rigorous proof of this fact for
subsequent work. Indeed, there are good reasons for this conjecture: First,
in the context of nonlinear optics \cite{Haus1990, Yeang1999} it is
supported by the fact that, in the limit of $t\rightarrow \infty $,
continuum modes completely disperse out, hence the orthogonality condition
definitely holds. Second, the orthogonality of the Goldstone and continuum
modes is built into the procedure of the construction of Bogoliubov
eigenstates \cite{Takahashi2015}.

Operators $\delta \hat{\psi}^{\dagger }$ and $\delta \hat{\psi}$ may be
interpreted as creation/annihilation operators of the quantum fluctuations.
To properly define the action of the operators, one has to specify the
nature of the vacuum state. The breather is initialized as a \textit{mother
soliton}, which defines the vacuum state of the
quantum-fluctuation operators around the breather. Below we address two
different physically relevant schemes for incorporating the vacuum state
into the Bogoliubov method.

\paragraph{The white-noise vacuum.} The most common approach to introduce
the vacuum state for the $\delta \hat{\psi}^{\dagger }$ and $\delta \hat{\psi%
}$ operators (in particular, in optics \cite{Haus1990, Lai1993, Haus1996,
Yeang1999, Haus2000}) is to consider one with fluctuations in the form of
uncorrelated random noise. Such a formulation is also adopted in atomic
physics \cite{Opanchuk2017}, and has the following interpretation: the
mother soliton is a Hartree product of non-interacting single-particle wave
functions, all having the shape of the mother soliton. Thus, only the product $%
\langle \delta \hat{\psi}(x,0)\delta \hat{\psi}^{\dagger }(x^{\prime
},0)\rangle =\delta (x-x^{\prime })$, where the averaging $\langle
...\rangle $ is taken over the vacuum state, defines nonzero correlations
(see \cite{supplement}). At $t=0$, quantum fluctuations of 8 parameters, $%
\Delta \hat{\chi}_{0}$, can be expressed in terms of overlaps of functions $%
f_{\chi }(x,t)$ as 
\begin{equation}
\langle \Delta \hat{\chi}_{0}^{2}\rangle \propto \int_{-\infty }^{+\infty }%
\mathrm{d}x\left\vert f_{\tilde{\chi}}(x,0)\right\vert ^{2}
\label{variance_eq}
\end{equation}%
(see \cite{supplement}), with 8 parameters combined in four
pairs ($\chi ,\tilde{\chi})$, \textit{viz}.,
($N$,$\Theta $), ($V$,$B$), ($n$,$\theta $), and ($v$,$b$). The relationships between them 
resemble canonical conjugation (up to constant factors). (See~\cite{supplement} for details.)
Derivatives of 
Gordon's solution and the overlap integrals were evaluated analytically
(using Wolfram Mathematica). The so evaluated fluctuations are presented in
Table~\ref{tab:init_variances}, where scales of the length and velocity are $%
\bar{x}=\hbar ^{2}/(mg)$ and $\bar{v}=g/\hbar $. For ${}^{7}$Li atoms with 
$m=7$ AMU, $\omega _{\perp }=254\times 2\pi $~Hz, and 
$a_{\mathrm{sc}}=-4a_{\mathrm{0}}$ ($a_{0}$ is the Bohr radius), 
we have $\bar{x}\approx 1.34$ cm
and $\bar{v}\approx 6.75\times 10^{-5}$ cm/s, while the breather's
oscillation period is $T_{\mathrm{br}}\approx 4\times 10^{6}/N^{2}$ s.

\begin{table}[h]
\begin{tabular}{|c|c|c|c|c|}
\hline
& Number & Phase & Velocity & Coordinate \\ \hline
Over. & $N$ & $\frac{105+ 11\pi ^{2}}{315N}$ & $\frac{N\bar{v}^{2}}{192}$ & $%
\frac{16\pi ^{2}\bar{x}^{2}}{3N^{3}}$ \\ 
Rel. & $N/5$ & $\frac{4(420+23\pi ^{2})}{315N}$ & $\frac{23N\bar{v}^{2}}{420}
$ & $\frac{256\pi ^{2}\bar{x}^{2}}{15N^{3}}$ \\ \hline
\end{tabular}%
\caption{Initial values of the quantum fluctuations $\langle \Delta \hat{%
\protect\chi}_{0}^{2}\rangle $ of the overall and relative parameters of the
breather, obtained for the \textit{white-noise} vacuum state. }
\label{tab:init_variances}
\end{table}
Next, we compare these uncertainty expressions with the standard
(Heisenberg's) ones: 
\begin{subequations}
\label{uncertainty-rels}
\begin{align}
& \langle \Delta \hat{N}_{0}^{2}\rangle \langle \Delta \hat{\Theta}%
_{0}^{2}\rangle \approx 0.678>0.25 \\
& N^{2}m^{2}\langle \Delta \hat{V}_{0}^{2}\rangle \langle \Delta \hat{B}%
_{0}^{2}\rangle /\hbar ^{2}\approx 0.274>0.25,  \label{VRuncert} \\
\label{uncertainty-ntheta}
&\langle \Delta \hat{\theta}_{0}^{2}\rangle  
\langle \Delta \hat{n}_{0}^{2}\rangle /4\approx 0.41 > 0.25, \\
& N^{2}(3m/16)^{2}\langle \Delta \hat{v}_{0}^{2}\rangle \langle \Delta \hat{b%
}_{0}^{2}\rangle /\hbar ^{2}\approx 0.3243 > 0.25,  \label{vb_uncertainty}
\end{align}%
where  the rightmost bound comes from the exact commutation relations between
$-\Theta$, $B$, $-\theta$, and $b$, and the corresponding ``momenta'' $\hbar N$,
$N m V$,  $\hbar n/2$, and $3N m v/16$ respectively, see \cite{supplement}.

Note that the
uncertainty value for the relative momentum, $3mNv/16$%
, and distance, $b$, is $\approx 20\%$ larger than that for COM
momentum-position pair. One can also evaluate averages of the cross-products
of the operators, using formulas similar to Eq. \eqref{variance_eq}, see
\cite{supplement}. Nonvanishing values 
\end{subequations}
\begin{equation}
\begin{aligned} \label{cross_products} \langle \Delta \hat N_0 \Delta \hat
\Theta_0 \rangle = i/2 ,\quad & \langle \Delta \hat B_0 \Delta \hat V_0
\rangle = i\hbar/(2 Nm)\\ \langle \Delta \hat n_0\Delta \hat \theta_0
\rangle = i ,\quad& \langle \Delta \hat b_0\Delta \hat v_0 \rangle = 8
i\hbar/(3 Nm) \end{aligned}
\end{equation}%
are purely imaginary due to properties of modes $f_{\bar{\chi}}$, and $%
\langle \Delta \hat{\chi}_{0}\Delta \hat{\chi}_{0}^{\prime }\rangle
=-\langle \Delta \hat{\chi}_{0}^{\prime }\Delta \hat{\chi}_{0}\rangle $ due
to the hermiticity.
Note that $\sqrt{\langle \Delta \hat{N}_{0}^{2}\rangle \langle \Delta \hat{%
\Theta}_{0}^{2}\rangle }\approx 0.82$, $\sqrt{\langle \Delta \hat{V}%
_{0}^{2}\rangle \langle \Delta \hat{B}_{0}^{2}\rangle }\approx 2.1\hbar
/(Nm) $, $\sqrt{\langle \Delta \hat{n}_{0}^{2}\rangle \langle \Delta \hat{%
\theta}_{0}^{2}\rangle }\approx 1.3$, and $\sqrt{\langle \Delta \hat{v}%
_{0}^{2}\rangle \langle \Delta \hat{b}_{0}^{2}\rangle }\approx 3\hbar /(Nm)$%
. Then, the cross term $\langle \Delta \hat{B}_{0}\Delta \hat{V}_{0}\rangle $
may be neglected, while others are non-negligible.

\paragraph{Contributions from mother-soliton's continuum fluctuations.}  The
predictions for fluctuations of the breather's parameters are significantly
different if field fluctuations of the mother (pre-quench) soliton are
included. In contrast to the white-noise vacuum case, we cannot keep only
one product of the fluctuating operators, $\delta \hat{\psi}(x)\delta \hat{%
\psi}^{\dagger }(x^{\prime })$, therefore the correlated-quantum-noise
vacuum leads to different expectation values. In turn, quantum fluctuations
of the fundamental mother soliton can be separated into discrete and
continuum parts~\cite{Lewenstein1996, Castin1998, Castin2009}. Further,
expectation values of the continuum creation/annihilation operator products
can be calculated using known exact expressions~\cite{Kaup1990, Castin2009}
for the Bogoliubov modes of the fundamental soliton (see \cite{supplement}).
Fluctuations of discrete parameters of the mother soliton are determined by
derivatives of the mean field with respect to these parameters. They
coincide with the breather's overall (COM) fluctuations, as the soliton's
and breather's mean fields are the same at $t=0$. Because fluctuations of
the discrete parameters of the mother soliton are decoupled from the
relative degrees of freedom of the breather, they do not affect the
corresponding variances. Uncertainties of the overall degrees of freedom of
the breather are determined by parameters of the experiment that creates the
mother soliton. Note also that, due to phase-diffusion effects~\cite%
{Lewenstein1996, Castin1998}, fluctuations of the discrete parameters of the
mother soliton depend on time between the creation of the soliton and the
application of the interaction quench to it. 
\begin{figure}[tbp]
\centering
\includegraphics[width=0.46\textwidth]{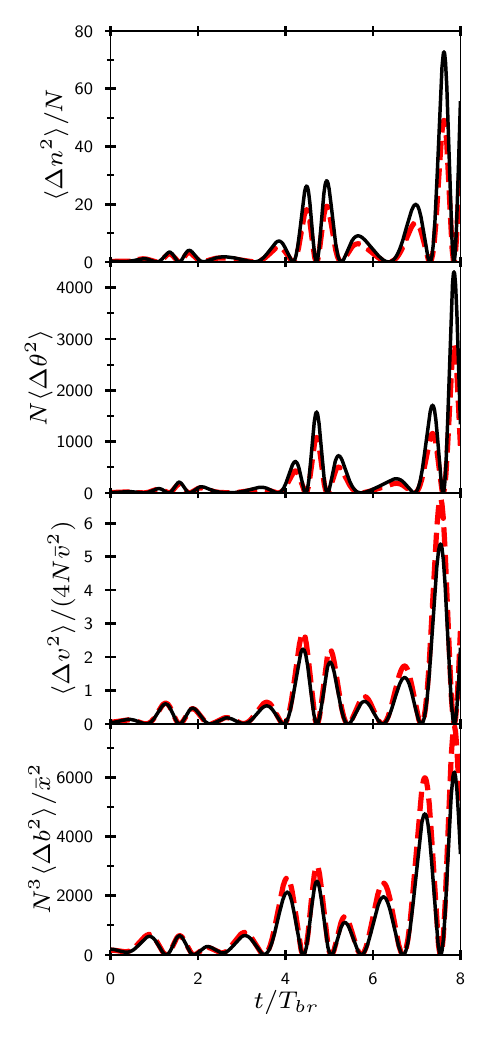} %
\caption{Variances of fluctuations of the relative parameters of the
breather, as a function of time (from top to bottom): the number of atoms $%
\langle \Delta \hat{n}^{2}(t)\rangle $, phase $\langle \Delta \hat{\protect%
\theta}^{2}(t)\rangle $, velocity $\langle \Delta \hat{v}^{2}(t)\rangle $,
and position $\langle \Delta \hat{b}^{2}(t)\rangle $, as found for the
white-noise vacuum state (red dashed lines) and pre-quench correlated
vacuum state (black solid lines).}
\label{evol_fluct_rel}
\end{figure}

In Table~\ref{tab:rel_variances} we compare initial variances of the
relative parameters for different vacuum states. Due to the complicated form
of the expressions, the variances for the correlated-noise vacuum were
evaluated numerically. The difference, while not being enormous, is evident
and it may manifest itself in the observable dynamics of the breather. The
cross-product averages are the same as for the
white-noise vacuum, see Eq. (\ref{cross_products}). 

The BA estimates for the relative velocity
variance obtained for small $N$ \cite{Yurovsky2017}, $0.035N\bar{v}^{2}$, is
within $20\%$ of the correlated-vacuum prediction. This conclusion
is an important result of the present work, as it demonstrates that the
crucially important characteristics of the fluctuational dynamics are close
for different vacuum states and
in the opposite limits of small and large $N$, thus revealing
\emph{universal features} of the dynamics, which should be amenable to
experimental observation. 

\begin{table}[h]
\begin{tabular}{|c|c|c|c|c|}
\hline
Noise & $\langle \Delta \hat n_0^2 \rangle$ & $\langle \Delta \hat
\theta_0^2 \rangle$ & $\langle \Delta \hat v_0^2 \rangle$ & $\langle\Delta
\hat b_0^2 \rangle$ \\ \hline
White & $0.2 N $ & $8.22/N $ & $0.0548 N\bar{v}^2$ & $168\bar{x}^2/N^3$ \\ 
Correlated & $0.3 N $ & $6.26/N $ & $0.0429 N \bar{v}^2$ & $198\bar{x}^2/N^3$
\\ \hline
\end{tabular}%
\caption{Initial quantum fluctuations of relative parameters of the
breather, for the white-noise and pre-quench correlated-vacuum states.}
\label{tab:rel_variances}
\end{table}

In Fig.~\ref{evol_fluct_rel} we display the evolution of the variances of
quantum operators of the relative parameters of the breather, and compare
the results for the white-noise and pre-quench correlated-noise vacuum
states. 

The splitting of the breather can be detected once the
constituent solitons are separated by a distance comparable to the
breather's width, which is $l_{\mathrm{br}}=8\hbar ^{2}/\left( mgN\right)
\approx 36~\mathrm{\mu }$m~\cite{Zakharov1972} under realistic experimental 
conditions ($N=3\times 10^{3}$ ${}^{7}$Li atoms 
with the parameters mentioned above).
Therefore, the time of dissociation due to quantum fluctuations,
$\tau =l_{\mathrm{br}}/\sqrt{\langle \Delta v_{0}^{2}\rangle }$, depends on the 
vacuum state (see Tab.~\ref{tab:rel_variances}), namely, 
$\tau _{\mathrm{white}}\approx 4.16$~s,
and $\tau _{\mathrm{corr}}\approx 4.7$~s. Thus, the inclusion of the
continuum fluctuations of the mother soliton increases the dissociation time
by more than a breather's period 
($\approx 0.22$ s).
Note that the BA estimate for small $N$\cite{Yurovsky2017}, 
$\approx 3$~s,  used a different
technical definition of the dissociation time; using the present definition,
the  BA yields  
$\tau _{\mathrm{BA}}\approx 5.18$~s. 
Eventually, the results again clearly corroborate the inference that the
fluctuational dynamics reveals \emph{universal features}, amenable to
experimental observation, in both limits of small and large $N$.

As noted above, the spontaneous dissociation
is forbidden in the integrable 1D axially uniform MF
model. The integrability maintains robustness of solitons and
``debris'', such as radiation or additional small-amplitude solitons, created in the experiment. The ``debris'' cannot bind into the breather, and
would disperse by themselves. In principle, dissociation
may be induced by integrability-breaking 3D effects,
decoherence, or an axial potential, which all are
unavoidable in the experiment. However, 3D MF calculations
\cite{Golde2018}, as well as analytical and numerical analyses
\cite{Pereira1979, Blow1985, Malomed1985, Dianov1986} of the decoherence, induced by the linear loss, do not
reveal any dissociation.
Besides, the relative motion of the constituent solitons is rather insensitive to long-scale potentials since 
linear potentials depend only on the COM coordinate and the quadratic ones cannot induce dissociation due to parity conservation. Calculations  \cite{Marchukov2019} demonstrated dissociation  due to a  narrow potential barrier (of the width  
$\delta x\ll l_{\mathrm{br}}$) above certain threshold, namely, the potential 
$\delta U$  does not induce dissociation if  
$\delta U\delta x\lesssim 10^{-4} g N$. This condition is not too strong in real experiments 
with large $N$. 
Thus, the dissociation into daughter-solitons can only be a result of quantum noise.
\paragraph{Conclusions.} The Bogoliubov linearization approach makes it possible
to estimate variances of the quantum fluctuations of the breather's discrete
parameters, including its COM and relative degrees of freedom. 
We consider two cases of the vacuum state: an easier, tractable uncorrelated
quantum noise, alias \textquotedblleft white noise\textquotedblright , and a
state with the correlated quantum noise, that takes into account quantum
fluctuations of the mother soliton. The comparison shows that the correlated
noise noticeably changes initial values and the evolution of the variances.
Disagreement between the relative-velocity variance for the correlated noise
and BA results \cite{Yurovsky2017}, obtained for the small number of atoms, 
$N\leq 23$, is $<20\%$. The present analysis produces variances of
other breather parameters as well. A fundamental observable that
quantum fluctuations can induce is dissociation of the breather. This effect
is essentially the same, irrespective of the choice of the noise pattern.
Namely, the dissociation time estimated for realistic experimental
parameters as $\tau _{\mathrm{corr}}\approx 4.7$~s for the correlated-noise
vacuum is about one breather period larger than for the uncorrelated
noise and closer to the BA estimate
 $\approx 5.18$ s for small $N$. The proximity of the basic results obtained
for small and large $N$ and for different vacuum states is a
strong indication that the quantum dynamics of breathers is dominated
by the universal features.  Thus, the results reveal the
feasibility of the observation of direct manifestations of quantum
fluctuations in macroscopic degrees of freedom---in particular, the relative
velocity of the two initially bound solitons. Note also that the proximity
of the uncertainty relation (\ref{vb_uncertainty}) to the lower limit of the
Heisenberg's position-momentum uncertainty relation indicates that the state
of the relative motion is \emph{probably} a macroscopically quantum one: if
it were spread over a large phase-space area, it would---while remaining
formally a pure state---become chaotic in the course of the subsequent
quantum evolution, while we see that it does not do that.

\begin{acknowledgments}
This work was jointly supported by the National Science Foundation through
grants No. PHY-1402249, No. PHY-1607221, and No. PHY-1912542 and the
Binational (US-Israel) Science Foundation through grant No. 2015616. The work at Rice
was supported by the Army Research Office Multidisciplinary University
Research Initiative (Grant No. W911NF-14-1-0003), the Office of Naval
Research, the NSF (Grant No. PHY-1707992), and the Welch Foundation (Grant
No. C-1133). The work at TAU was supported by the Israel Science Foundation (grant No. 1287/17) and by the Tel Aviv University -
Swinburne University of Technology Research Collaboration award. OVM acknowledges the support from the German Aeronautics and Space Administration (DLR) through Grant No. 50 WM 1957.
\end{acknowledgments}


%%%%%%%%%%%%%%%%%%%%%%%%%%%%%%%%%%%%%%%%%%%%%%%%%%%%%%%%%%%%%%%%%%%%%%%%%%%%%%%%%%%%%%
\pagebreak
\setcounter{equation}{0}
\onecolumngrid
\counterwithin*{equation}{section}
\begin{center}
\bf{Supplemental material for: Quantum fluctuations of the center-of-mass and relative parameters of NLS breather} \end{center}
\begin{center}
O. V. Marchukov, B. A. Malomed,  V. Dunjko, J. Ruhl, M. Olshanii, R. G. Hulet, and V. A. Yurovsky
\end{center}

\newcommand*{\linqnlse}
{(2)}
\newcommand*{\quancorr}
{(3)}
\newcommand*{\crossproducts}
{(6)}
\newcommand*{\initvariances}
{I}
\newcommand*{\cross}
{II}
\newcommand*{\mainNLSE}
{(1)}
\newcommand*{\fieldexpect}
{(4)}
\newcommand*{\uncertaintyrels}
{(5)}

\renewcommand{\theequation}{S-\arabic{equation}} \renewcommand{\thefigure}{S\arabic{figure}}

Numbers of equations and figures in the Supplemental material
start with S. References to equations and figures in the Letter do not
contain S.

\section{The Gordon two-soliton solution}\label{secGordon}

The mean field $\Psi _{0}(x,t)$ is the solution of the classical NLS (GP)
equation, 
\begin{equation}
i\frac{\partial \Psi _{0}}{\partial t}=-\frac{1}{2}\frac{\partial ^{2}\Psi
_{0}}{\partial x^{2}}-N|\Psi _{0}|^{2}\Psi _{0},  \label{gpe}
\end{equation}%
where the coordinates and time,  respectively, are measured in units of 
\begin{equation}\label{unitsxbartbar}
\bar{x}=\hbar^{2}/(mg), \quad  \bar{t}=\hbar ^{3}/(mg^{2}).
\end{equation} 
For two solitons with the velocities $v_1$ and $v_2$ and 
initial positions  $b_1$ and $b_2$
containing $N_{1}$ and $N_{2}$ atoms, respectively, the solution was obtained in 
Ref. \cite{Gordon1983}, 
%\begin{widetext}
\begin{align}
& \Psi _{0}(x,t)=\frac{\sqrt{N}}{2}\left( \Phi _{+}(x-B-Vt,t)+\Phi
_{-}(x-B-Vt,t)\right) \exp {\left( i\phi (x-B-Vt,t)+iVx-iV^{2}t/2+i\Theta
\right) ,}  \notag \\
& \Phi _{\pm }(x,t)=e^{\pm i\varphi }\frac{(1\pm n/N)\left( \frac{Nn\pm
4v^{2}}{N^{2}}\right) \cosh (\frac{Nx}{4}\mp z)-i((n/N)^{2}-1)\frac{2v}{N}%
\sinh (\frac{Nx}{4}\mp z)}{\left( 1-(n/N)^{2}\right) \cos (2\varphi )+\left( 
\frac{n^{2}+4v^{2}}{N^{2}}\right) \cosh (\frac{N}{2}x)+\left( \frac{4v^{2}}{%
N^{2}}+1\right) \cosh (2z)},  \label{Gordon}
\end{align}%
%
%
%\end{widetext}
where $\phi (x, t)=\frac{1}{2}\left(1 + \frac{n^2}{N^2}\right) \left( \frac{N^{2}-4v^{2}}{16}\right) t   -\frac{n}{2N}v x$, 
$z(x,t)=\frac{N}{4}\left( nx/N - \frac{1}{2}\left( 1-(n/N)^{2}\right) (b+tv)\right) $, and 
$\varphi(x,t) =\frac{n}{4N}v^{2}t+n\frac{N}{16}t+\frac{1}{2}vx +\frac{\theta }{2}$. This
solution depends on 8 parameters, namely, the total number of atoms $%
N=N_{1}+N_{2}$, phase $\Theta $, COM velocity $V$, initial COM coordinate $B$%
, relative velocity of the constituent solitons $v$, initial distance
between the solitons $b$, relative phase difference $\theta $, and the
atomic-number difference, $n=(N_{2}-N_{1})$. The velocities are measured in
units of $\bar{v}=g/\hbar $.
The solution is normalized so that $%
\int_{-\infty }^{+\infty }\mathrm{d}x|\Psi _{0}(x,t)|^{2}=1$. 
We use Galilean invariance  and set $B=V=0$, like in previous applications
\cite{Haus1985, Haus1990, Lai1993, Yeang1999} of the linearization method to the COM motion.
These parameters are kept in Eq. (\ref{Gordon}) since we use its derivatives over them.
%--------------------------------------

\section{The relation between Hermitian operators $\Delta \hat{\protect\chi}%
_{0}$ and fluctuations of parameters $\protect\chi $ of the mean-field
solution}

\label{a:fluct_parameters} To derive the relation, we use the fact that the
ensemble average of the mean-field-solution density matrix gives the same
result as the expectation value of the density operator constructed from the
quantum field: 
\begin{equation}
\overline{\Psi _{0}^{\ast }(x,t;\chi )\Psi (x,t;\chi )}=\langle \hat{\Psi}%
^{\dagger }(x,t)\hat{\Psi}(x,t)\rangle 
\end{equation}%
To calculate the density, we need to know mean field $\tilde{\Psi}_{0}$ that
takes into account fluctuations according to the Hartree-Fock-Bogoliubov
equation~\cite{Griffin1996, Andersen2004}: 
\begin{equation}
i\frac{\partial \tilde{\Psi}_{0}}{\partial t}=-\frac{1}{2}\frac{\partial ^{2}%
\tilde{\Psi}_{0}}{\partial x^{2}}-N|\tilde{\Psi}_{0}|^{2}\tilde{\Psi}%
_{0}-2\left\langle \delta \hat{\psi}^{\dagger }\delta \hat{\psi}%
\right\rangle \tilde{\Psi}_{0}-\left\langle \delta \hat{\psi}\delta \hat{\psi%
}\right\rangle \tilde{\Psi}_{0}^{\ast }.
\end{equation}%
It can be approximated as $\tilde{\Psi}_{0}\approx \Psi _{0}+\Psi _{2}$,
where $\Psi _{0}$ is the solution of NLSE \eqref{gpe}, and $\Psi _{2}$ is
the correction which satisfies the linear driven equation, 
\begin{equation}
i\frac{\partial \Psi _{2}}{\partial t}=-\frac{1}{2}\frac{\partial ^{2}\Psi
_{2}}{\partial x^{2}}-2N|\Psi _{0}|^{2}\Psi _{2}-N\Psi _{0}^{2}\Psi
_{2}^{\ast }-\sum_{\chi ,\chi ^{\prime }}\left( 2f_{\chi }^{\ast }\Psi
_{0}+f_{\chi }\Psi _{0}^{\ast }\right) f_{\chi ^{\prime }}\left\langle
\Delta \hat{\chi}_{0}\Delta \hat{\chi}_{0}^{\prime }\right\rangle ,
\label{Psi2eq}
\end{equation}%
where we use the discrete-mode part of expansion {\quancorr}, 
\begin{equation}
\delta \hat{\psi}=\sum_{\chi }f_{\chi }\Delta \hat{\chi}_{0},
\label{quancorrd}
\end{equation}%
with $f_{\chi }(x,t)=\partial (\sqrt{N}\Psi _{0})/\partial \chi $. Thus, $%
\Psi _{2}$ can be expressed as 
\begin{equation}
\Psi _{2}=\frac{1}{2N}\sum_{\chi ,\chi ^{\prime }}\frac{\partial ^{2}}{%
\partial \chi \partial \chi ^{\prime }}\Psi _{0}\left\langle \Delta \hat{%
\chi _{0}}\Delta \hat{\chi}_{0}^{\prime }\right\rangle ,
\end{equation}%
as the second derivative of $\Psi _{0}$ satisfies the differential equation
with the same homogeneous part as in Eq. (\ref{Psi2eq}). Then, using the
expansion of the field operator, $\hat{\Psi}(x,t)=\sqrt{N}\tilde{\Psi}%
_{0}(x,t)+\delta \hat{\psi}(x,t)$, we can calculate the density 
\begin{align}
\left\langle \hat{\Psi}^{\dagger }\hat{\Psi}\right\rangle & \approx N\Psi
_{0}^{\ast }\Psi _{0}+\frac{1}{2}N\sum_{\chi ,\chi ^{\prime }}\left( \Psi
_{0}^{\ast }\frac{\partial ^{2}}{\partial \chi \partial \chi ^{\prime }}\Psi
_{0}+\Psi _{0}\frac{\partial ^{2}}{\partial \chi \partial \chi ^{\prime }}%
\Psi _{0}^{\ast }+2\frac{\partial \Psi _{0}^{\ast }}{\partial \chi }\frac{%
\partial \Psi _{0}}{\partial \chi ^{\prime }}\right) \left\langle \Delta 
\hat{\chi}_{0}\Delta \hat{\chi}_{0}^{\prime }\right\rangle   \notag \\
& =N\Psi _{0}^{\ast }\Psi _{0}+\frac{1}{2}N\sum_{\chi ,\chi ^{\prime }}\frac{%
\partial ^{2}}{\partial \chi \partial \chi ^{\prime }}\left( \Psi _{0}^{\ast
}\Psi _{0}\right) \left\langle \Delta \hat{\chi}_{0}\Delta \hat{\chi}%
_{0}^{\prime }\right\rangle .
\end{align}%
On the other hand, we calculate the ensemble average of the classical field
solution \eqref{Gordon}, that depends on fluctuating parameters $\chi $,
using the Taylor expansion: 
\begin{equation}
\Psi _{0}(x,t;\{\chi\} )=\Psi _{0}(x,t;\{\chi _{0}\})
+\sum_{\chi }\left( \frac{\partial \Psi _{0}}{\partial \chi }\right) _{\chi =\chi _{0}}\delta \chi 
+\frac{1}{2}\sum_{\chi ,\chi ^{\prime }}
\left( \frac{\partial ^{2}}{\partial\chi \partial \chi ^{\prime }}\Psi _{0}\right) _{\chi =\chi _{0}}
\delta \chi\delta \chi ^{\prime }+\dots ,
\end{equation}%
where $\delta \chi =\chi -\chi _{0}$. Then, the mean-field density $N\Psi
_{0}^{\ast }\Psi _{0}$, averaged over classical fluctuations of parameters $%
\delta \chi $ is (linear terms vanish here) 
\begin{equation}
N\overline{\Psi _{0}^{\ast }\Psi _{0}}\approx N\Psi _{0}^{\ast }\Psi _{0}+%
\frac{1}{2}N\sum_{\chi ,\chi ^{\prime }}\frac{\partial ^{2}}{\partial \chi
\partial \chi ^{\prime }}\left( \Psi _{0}^{\ast }\Psi _{0}\right) \overline{%
\delta \hat{\chi}\delta \hat{\chi}^{\prime }}.
\end{equation}%
Thus the quantum and classical fluctuations lead to the same density
corrections, provided that 
\begin{equation}
\overline{\delta \chi \delta \chi ^{\prime }}=\left\langle \Delta \hat{\chi}%
_{0}\Delta \hat{\chi}_{0}^{\prime }\right\rangle .
\end{equation}

\section{Calculation of quantum fluctuations of the breather's parameters 
\label{a:fluctuations}}

Functions $f_{\chi }(x,t)=\partial (\sqrt{N}\Psi _{0})/\partial \chi $,
which are derivatives of the solution of GPE \eqref{gpe}, multiplied by $%
\sqrt{N}$, are c-number solutions of the linearized NLSE \linqnlse. One may
introduce a conservation relation~\cite{Haus1985, Haus1990}, 
\begin{equation}
\frac{\partial }{\partial t}\int_{-\infty }^{+\infty }\left[ (\mathrm{Re}%
f_{\chi })(\mathrm{Re}\bar{f}_{\chi })+(\mathrm{Im}f_{\chi })(\mathrm{Im}%
\bar{f}_{\chi })\right] \mathrm{d}x=0,
\end{equation}%
where $\bar{f}_{\chi }$ is the solution of the equation adjoint to Eq.~%
\linqnlse,%
\begin{equation}
i\frac{\partial \bar{f}_{\chi }}{\partial t}=-\frac{1}{2}\frac{\partial ^{2}%
\bar{f}_{\chi }}{\partial x^{2}}-2|\Psi _{0}|^{2}\bar{f}_{\chi }+\Psi
_{0}^{2}\bar{f}_{\chi }^{\ast }.  \label{adj-lin-qnlse}
\end{equation}%
Solution $\bar{f}_{\chi }$ is related to $f_{\chi }$ as $\bar{f}_{\chi
}(x,t)=if_{\chi }(x,t)$. These adjoint functions fulfill orthogonality
conditions 
\begin{equation}
C_{\chi \xi }=\prec \bar{f}_{\chi }|f_{\xi }\succ ,  \label{orthcon}
\end{equation}%
with the quasi-inner product defined as 
\begin{equation}
\prec \bar{f}_{\chi }|f_{\xi }\succ =\int_{-\infty }^{+\infty }\mathrm{d}x%
\left[ (\mathrm{Re}~\bar{f}_{\chi })(\mathrm{Re}~f_{\xi })+(\mathrm{Im}~\bar{%
f}_{\chi })(\mathrm{Im}~f_{\xi })\right]  = \frac{i}{2} \int_{-\infty }^{+\infty }\mathrm{d}x
\left [ f_\chi f_\xi^\ast - f_\chi^\ast f_\xi \right ],
\label{inner}
\end{equation}%
for derivatives $f_{\chi }$ and $f_{\xi }$. The derivatives of the
two-soliton Gordon solution \eqref{Gordon} with respect to 8 parameters $N$, $%
\Theta $, $V$, $B$, $n$, $\theta $, $v$, $b$, as well as integrals in Eq. %
\eqref{inner}, can be calculated analytically. The only non-zero quasi-inner
products of the derivatives at $t=0$ are 
\begin{subequations}
\label{orth_com}
\begin{align}
& C_{N\Theta }=-C_{\Theta N}=\frac{1}{2}, \\
& C_{VB}=-C_{BV}=\frac{N}{2}, \\
& C_{n\theta }=-C_{\theta n}=1/4, \\
& C_{vb}=-C_{bv}=\frac{3N}{32},
\end{align}%
with all other quasi-inner products vanishing. Thus, the parameters can be
combined into four pairs, namely, ($N$, $\Theta $), ($V$, $%
B$), ($n$, $\theta $), and ($v$, $b$). Only the inner products corresponding
to the ``paired'' parameters $\chi $ and $\tilde{\chi}$ do not vanish, having the property $C_{\chi \tilde{\chi}}=-C_{\tilde{\chi}\chi }$.
As was mentioned in the main text the parameters pairs resemble the canonically conjugated variables but 
the connection is not trivial (see Sec.~\ref{a:canonical_structure} of this Supplemental Material).
Moreover, it is tempting to conjecture a connection between the coefficients 
$C_{\chi \xi }$ and the Poisson brackets of canonical variables and this connection is a subject of future research.

Using the quantum-correction expansion \eqref{quancorrd}, we find the expression for
the initial ($t=0$) quantum fluctuations of the parameters: 
\end{subequations}
\begin{equation}
\Delta \hat{\chi}_{0}=C_{\tilde{\chi}\chi }^{-1}\prec \bar{f}_{\tilde{\chi}%
}(x,0)|\delta \hat{\psi}(x,0)\succ ,  \label{init-fluc}
\end{equation}%
where the real\ and imaginary\ parts of the operators in the quasi-inner
products \eqref{inner} are defined as $\mathrm{Re}~\delta \hat{\psi}=(\delta 
\hat{\psi}+\delta \hat{\psi}^{\dag })/2$, $\mathrm{Im}~\delta \hat{\psi}%
=-i(\delta \hat{\psi}-\delta \hat{\psi}^{\dag })/2$. Similarly, quantum
fluctuations may be defined for $t>0$ as \cite{Haus1990, Yeang1999} 
\begin{equation}
\Delta \hat{\chi}(t)=C_{\tilde{\chi}\chi }^{-1}\prec \bar{f}_{\tilde{\chi}%
}(x,0)|e^{-7iN^{2}t/128}\delta \hat{\psi}(x,t)\succ ,  \label{fluc_evol}
\end{equation}%
where the exponential factor cancels the mean-field phase shift. Using
expansion \eqref{quancorrd} of operator $\hat{\psi}(x,t)$, we can express
the quantum fluctuations at $t>0$ in terms of the initial quantum
fluctuations as 
\begin{equation}
\Delta \hat{\chi}(t)=\sum_{\xi }M_{\chi \xi }\Delta \hat{\xi}_{0},
\label{fluc_mat_eq}
\end{equation}%
where 
\begin{equation}
M_{\chi \xi }=C_{\tilde{\chi}\chi }^{-1}\prec \bar{f}_{\tilde{\chi}%
}(x,0)|e^{-7iN^{2}t/128}f_{\xi }(x,t)\succ ,  \label{Mchixi}
\end{equation}%
while the integrals here are calculated numerically.

Substituting the definition of the quasi-inner product \eqref{inner} in Eq. %
\eqref{init-fluc}, we can express the quantum-fluctuation operators in an
explicit form, 
\begin{equation}  \label{Deltachi0}
\Delta \hat{\chi}_{0}=\frac{i}{2C_{\tilde{\chi}\chi }}\int_{-\infty
}^{\infty }\mathrm{d}x\left( f_{\tilde{\chi}}(x,0)\delta \hat{\psi}^{\dagger
}(x,0)-f_{\tilde{\chi}}^{\ast }(x,0)\delta \hat{\psi}(x,0)\right) .
\end{equation}%
Note that operators $\Delta \hat{\chi}_{0}$ are Hermitian, i.e., $\Delta 
\hat{\chi}_{0}^{\dagger }=\Delta \hat{\chi}_{0}$.
This gives the explicit form of Eq.~\fieldexpect of the main text
\begin{equation}
    \langle \Delta \hat{\chi}_{0}^{2}\rangle = \frac{1}{4 C_{\bar \chi \chi}^2} \int_{-\infty }^{+\infty }%
\mathrm{d}x\left\vert f_{\tilde{\chi}}(x,0)\right\vert ^{2}.
\end{equation}

\section{Calculation of expectation values in the white-noise vacuum \label%
{a:white-noise vacuum}}

We find variances of the quantum fluctuations by using the \textquotedblleft
white noise\textquotedblright\ vacuum states where the expectation values of
the fluctuation creation and annihilation operators $\delta \hat{\psi}%
^{\dagger }(x,0)$ and $\delta \hat{\psi}(x,0)$ are given by 
\begin{subequations}
\label{fieldexpect}
\begin{align}
& \langle \delta \hat{\psi}(x,0)\delta \hat{\psi}(x^{\prime },0)\rangle
=\langle \hat{\delta}\psi ^{\dagger }(x,0)\delta \hat{\psi}^{\dagger
}(x^{\prime },0)\rangle =0, \\
& \langle \delta \hat{\psi}^{\dagger }(x,0)\delta \hat{\psi}(x^{\prime
},0)\rangle =0, \\
& \langle \delta \hat{\psi}(x,0)\delta \hat{\psi}^{\dagger }(x^{\prime
},0)\rangle =\delta (x-x^{\prime }).
\end{align}%
Then Eq. \eqref{Deltachi0} leads to the average over vacuum of the products
of the fluctuation operators, 
\end{subequations}
\begin{equation}
\langle \Delta \hat{\chi}_{0}\Delta \hat{\xi}_{0}\rangle =\frac{1}{4C_{%
\tilde{\chi}\chi }C_{\tilde{\xi}\xi }}\int_{-\infty }^{\infty }\mathrm{d}xf_{%
\tilde{\chi}}^{\ast }(x,0)f_{\tilde{\xi}}(x,0).  \label{aver_fl_product}
\end{equation}%
Analytical integration leads then to the initial values of the quantum
fluctuation variances of the breather's parameters, which are given by Table~%
\initvariances~and Eq.~\crossproducts.

Finally, one can show that the evolution of quantum fluctuation variances
can be calculated as 
\begin{equation}
\langle \Delta \hat{\chi}^{2}(t)\rangle =\sum_{\xi }|M_{\chi \xi
}|^{2}\langle \Delta \hat{\xi}_{0}^{2}\rangle   \label{var_mat_eq}
\end{equation}%
[see Eq. (\ref{Mchixi})]. %--------------------------------------

\section{Calculation of expectation values in the correlated vacuum \label%
{a:corrected vacuum}}

The quantum field of the mother soliton at $t=0$ is taken as $\hat{\Phi}(x)=%
\sqrt{N}\Phi _{0}(x)+\delta \hat{\psi}_{\mathrm{cont}}+\delta \hat{\psi}_{%
\mathrm{discr}}$. The soliton solution is 
\begin{equation}
\Phi _{0}(x)=\frac{1}{2\hbar }\sqrt{m|g_{0}|N}\mathrm{sech}\left( \frac{%
m|g_{0}|N}{2\hbar ^{2}}(x-B)\right) \exp \left (i\frac{m}{\hbar }Vx+i\Theta\right ).
\end{equation}%
It depends on the same overall parameters $N$, $\Theta $, $V$, and $B$ as
the two-soliton solution (\ref{Gordon}). 
As the mother soliton is a pre-quench solution, we here take $|g_{0}|=g/4$ 
and, hereafter, do not use units defined by Eq. \eqref{unitsxbartbar}.
As it was mentioned above, we use Galilean invariance and set $B=V=0$.
The continuum part can be expressed as~\cite{Kaup1990, Castin2009}
\begin{equation}
\delta \hat{\psi}_{\mathrm{cont}}=\int_{-\infty }^{+\infty }\frac{\mathrm{d}k%
}{2\pi }\left( U_{k}(x)\hat{b}_{k}+V_{k}^{\ast }(x)\hat{b}_{k}^{\dagger
}\right) ,
\end{equation}%
with 
\begin{subequations}
\begin{equation}
U_{k}(x)=\frac{1+(K^{2}-1)\cosh ^{2}{X}+2iK\sinh {X}\cosh {X}}{%
(K-i)^{2}\cosh ^{2}{X}}e^{iKX},
\end{equation}%
\begin{equation}
V_{k}(x)=\frac{1}{(K-i)^{2}\cosh ^{2}{X}}e^{iKX},
\end{equation}%
where 
\end{subequations}
\begin{equation}
X=\frac{m|g_{0}|N}{2\hbar ^{2}}x,\quad K=\frac{2\hbar ^{2}}{m|g_{0}|N}k,
\end{equation}%
and operators $\hat{b}_{k}$ and $\hat{b}_{k}^{\dagger }$ obey the standard
bosonic commutation relations, $[\hat{b}_{k},\hat{b}_{k^{\prime }}^{\dagger
}]=2\pi \delta (k-k^{\prime })$. The expectation values of the products of
the mother-soliton's continuum-fluctuation operators can be calculated
analytically. As $\delta \hat{\psi}(x,0)=\delta \hat{\psi}_{\mathrm{cont}%
}+\delta \hat{\psi}_{\mathrm{discr}}$, for the breather's fluctuations we
obtain 
\begin{subequations}
\label{bogoliubov_corrections}
\begin{align}
& \langle \delta \hat{\psi}(x,0)\delta \hat{\psi}(x^{\prime },0)\rangle =
\frac{m|g_{0}|N}{4\pi \hbar ^{2} }
\left( -\frac{\pi}{2} \text{sech}^{2}(X)\text{sech}^{2}(X^{\prime })
e^{-\left\vert X-X^{\prime }\right\vert }(\cosh(2X)\left\vert X-X^{\prime }\right\vert 
+(X-X^{\prime })\sinh (2X)-1)\right) 
  \notag \\
& +\langle \delta \hat{\psi}_{\mathrm{discr}}(x)
\delta \hat{\psi}_{\mathrm{discr}}(x^{\prime })\rangle , 
\\
& \langle \delta \hat{\psi}^{\dagger }(x,0)
\delta \hat{\psi}^{\dagger}(x^{\prime },0)\rangle =
\langle \delta \hat{\psi}(x^{\prime },0)\delta \hat{\psi}(x,0)\rangle ^{\ast }, 
\\
& \langle \delta \hat{\psi}^{\dagger }(x,0)\delta \hat{\psi}(x^{\prime},0)\rangle =\frac{m|g_{0}|N}{4\pi \hbar ^{2} }\left( \frac{\pi }{2}\text{sech}^{2}(X)
\text{sech}^{2}(X^{\prime })e^{-\left\vert X-X^{\prime }\right\vert }
(\left\vert X-X^{\prime }\right\vert +1)\right) 
+\langle \delta \hat{\psi}_{\mathrm{discr}}^{\dagger }(x)
\delta \hat{\psi}_{\mathrm{discr}}(x^{\prime })\rangle , 
\\
& \langle \delta \hat{\psi}(x,0)\delta \hat{\psi}^{\dagger }(x^{\prime},0)\rangle =
\delta (x-x^{\prime })
+\langle \delta \hat{\psi}^{\dagger}(x^{\prime },0)\delta \hat{\psi}(x,0)\rangle .
\end{align}
\end{subequations}
As in this case all of the products of fluctuation operators $\delta \hat{%
\psi}$ and $\delta \hat{\psi}^{\dagger }$ yield nonzero contributions,
variances of parameter operators can be expressed as 
\begin{equation}
\langle \Delta \hat{\chi}_{0}^{2}\rangle =-\frac{1}{2|C_{\tilde{\chi}\chi
}|^{2}}\mathrm{Re}\int_{-\infty }^{+\infty }\mathrm{d}x\mathrm{d}x^{\prime
}\left( f_{\tilde{\chi}}(x,0)f_{\tilde{\chi}}(x^{\prime },0)\langle \delta 
\hat{\psi}^{\dagger }(x,0)\delta \hat{\psi}^{\dagger }(x^{\prime },0)\rangle
-f_{\tilde{\chi}}(x,0)f_{\tilde{\chi}}^{\ast }(x^{\prime },0)\langle \delta 
\hat{\psi}^{\dagger }(x,0)\delta \hat{\psi}(x^{\prime },0)\rangle \right) .
\label{full_variance}
\end{equation}%
The fluctuations of the mother-soliton's discrete parameters are
represented, as in Eq. \eqref{quancorrd}, in terms of derivatives of the soliton's
mean field: 
\begin{equation}
\delta \hat{\psi}_{\mathrm{discr}}(x)=\sum_{\xi \in \{N,\Theta ,V,B\}}\frac{%
\partial \sqrt{N}\Phi _{0}(x)}{\partial \xi }\Delta \hat{\xi}_{0}=\sum_{\xi
\in \{N,\Theta ,V,B\}}f_{\xi }(x,0)\Delta \hat{\xi}_{0}.
\end{equation}%
where the last equality is a consequence of matching of the mean fields
before and after the quench, $\Phi _{0}(x)=\Psi _{0}(x,0)$. Therefore, the
contribution of $\delta \hat{\psi}_{\mathrm{discr}}(x)$ to variances (\ref%
{full_variance}) of the relative parameters becomes proportional to the
inner products $C_{\chi \xi }$ [see Eq. (\ref{orthcon})] with $\xi \in
\{N,\Theta ,V,B\}$ and $\chi \in \{n,\theta ,v,b\}$.  But $C_{\chi \xi }=0$
[see Eq. (\ref{orth_com})], therefore fluctuations of the discrete parameters of the mother soliton
do not contribute to variances (\ref{bogoliubov_corrections}) of the
breather's relative parameters. Equations (\ref{bogoliubov_corrections}) and
(\ref{full_variance}) are used to calculate variances for the correlated
vacuum in the main text.

\newcommand{\FadTak}{}
\newcommand{\Gord}{\text{G}}
\renewcommand{\Re}{\operatorname{Re}}
\renewcommand{\Im}{\operatorname{Im}}
\newcommand{\FTA}[1]{{\ensuremath{A^{\FadTak}_{#1}}}}
\newcommand{\FTv}[1]{{\ensuremath{v^{\FadTak}_{#1}}}}
\newcommand{\FTxz}[1]{{\ensuremath{x_{0\,#1}^{\FadTak}}}}
\newcommand{\FTphiz}[1]{{\ensuremath{\phi_{0\,#1}^{\FadTak}}}}
\newcommand{\FTphiTz}[1]{{\ensuremath{\widetilde{\phi}{}_{0\,#1}^{\FadTak}}}}
\newcommand{\GA}[1]{\ensuremath{A^{\Gord}_{#1}}}
\newcommand{\Gv}[1]{\ensuremath{v^{\Gord}_{#1}}}
\newcommand{\Gxz}[1]{{\ensuremath{x_{0\,#1}^{\Gord}}}}
\newcommand{\Gphiz}[1]{{\ensuremath{\phi_{0\,#1}^{\Gord}}}}
\newcommand{\cFTp}[1]{{\ensuremath{p_{0,#1}^{\FadTak}}}}
\newcommand{\cFTq}[1]{{\ensuremath{q_{0,#1}^{\FadTak}}}}
\newcommand{\cFTrho}[1]{{\ensuremath{{{\varrho_{0,#1}^{\FadTak}}}}}}
\newcommand{\cFTphi}[1]{{\ensuremath{{{\varphi_{0,#1}^{\FadTak}}}}}}

\newcommand{\ctFTp}[1]{{\ensuremath{p_{#1}^{\FadTak}}}}
\newcommand{\ctFTq}[1]{{\ensuremath{q_{#1}^{\FadTak}}}}
\newcommand{\ctFTrho}[1]{{\ensuremath{{{\varrho_{#1}^{\FadTak}}}}}}
\newcommand{\ctFTphi}[1]{{\ensuremath{{{\varphi_{#1}^{\FadTak}}}}}}

\section{Canonical structure \label{a:canonical_structure}}
The canonical structure of the NLS equation is elaborated in \cite{takhtajan_book2007}.
Assuming a purely solitonic state consisting of $M$ solitons, there are four canonical variables per soliton, labeled $\ctFTq{j}$, $\ctFTphi{j}$,
$\ctFTp{j}$, and $\ctFTrho{j}$ for the $j$th soliton. 
In the notation of Ref.~\cite{takhtajan_book2007}, the Hamiltonian is
$H=\frac{1}{4}\sum_{j=1}^{M}\left(\ctFTrho{j} \ctFTp{j}^{2}-\frac{1}{3}\ctFTrho{j}^{3}\right)$
(without loss of generality, here we have set the parameter $\varkappa$ in \cite{takhtajan_book2007} to $-1$). 
The values of these canonical variables at $t=0$ are yet another set of canonical variables, $\cFTq{j}$, $\cFTphi{j}$,
$\cFTp{j}$, and $\cFTrho{j}$, which will be more convenient here. They satisfy the canonical relations 
\begin{align}
 \left\{\cFTq{j},\,\cFTp{k}\right\}&=\delta_{j\,k} && \left\{\cFTphi{j},\,\cFTrho{k}\right\}=\delta_{j\,k}\,,
\end{align}
while all other Poisson brackets are zero; see Eqs.~(7.13) and (7.14) on p.~231 of \cite{takhtajan_book2007}. 
 Here the Poisson bracket is defined as
\begin{equation}
 \left\{f,\,g\right\}=\sum_{j=1}^{M}\left[\left(\frac{\partial f}{\partial \cFTq{j}}\frac{\partial g}{\partial \cFTp{j}}-\frac{\partial g}{\partial \cFTq{j}}\frac{\partial f}{\partial \cFTp{j}}\right)+\left(\frac{\partial f}{\partial \cFTphi{j}}\frac{\partial g}{\partial \cFTrho{j}}-\frac{\partial g}{\partial \cFTphi{j}}\frac{\partial f}{\partial \cFTrho{j}}\right)\right]\,.
 \label{PoissonBracket}
\end{equation}

In order to connect the parametrization of Eq.~(\ref{Gordon}) (i.e.\ the relative and center-of-mass parameters $N$, $\Theta$, $V$, $B$, $v$, $b$, $\theta$, and $n$) to the canonical variables, as intermediaries we use the individual soliton parameters from Eqs.~(4) and (5) of \cite{Gordon1983}. For each constituent soliton $j$,  these are $A_{j}$, $v_{j}$, $x_{j0}$ and $\phi_{j0}$, where the $A_{j}$ is half the norm of the $j$th soliton, $A_{j}=N_{j}/2$. For convenience, we will relabel $x_{j0}$ and $\phi_{j0}$ by $b_{j}$ and $\theta_{j}$.

Most connections between the two sets of parameters are just as one would expect: $V=(N_{1}v_{1}+N_{2}v_{2})/(N_{1}+N_{2})$, $v=v_{2}-v_{1}$, $B=(N_{1}b_{1}+N_{2}b_{2})/(N_{1}+N_{2})$, $b=b_{2}-b_{1}$, $N=N_{1}+N_{2}$ and $n=N_{2}-N_{1}$. But two are nontrivial:
\begin{align}
 \theta&=\theta_{2}-\theta_{1} +\pi+\frac{n}{N}b v-b V  & \text{and} &&
 \Theta&=\frac{\theta_{1}+\theta_{2}}{2}-\frac{\pi}{2}
 -\frac{1}{4}\left(1+\frac{n^{2}}{N^{2}}\right)b v - B V + \frac{n}{2N}b V\,.
\end{align}

On the other hand, the individual soliton parameters can be related to the canonical variables. Here are the connections, where we have also inserted $\bar{x}$ and $\hbar$ as appropriate to restore the full dimensionalities:
\begin{align}
  \begin{aligned}
 b_{j}&=\frac{2\cFTq{j}\bar{x}\hbar}{\cFTrho{j}}  \\
 \theta_{j}&=-\cFTphi{j}+\frac{\cFTq{j}\cFTp{j}}{\cFTrho{j}} 
 \end{aligned}
 &&
  \begin{aligned}
  v_{j}&=\frac{\cFTp{j}}{2m\bar{x}}\\
   N_{j}&=\frac{\cFTrho{j}}{\hbar} \,.
 \end{aligned}
 \label{SIvariables}
\end{align}

In the main text, we  calculated uncertainty relations \uncertaintyrels\ for the four pairs of variables:
\begin{subequations}
\begin{align}
B = \frac{N_{1}b_{1} + N_{2}b_{2}}{N_{1}+N_{2}} & \qquad N m V= N m \frac{N_{1}v_{1} + N_{2}v_{2}}{N_{1}+N_{2}}
\\
b= b_{2}-b_{1} & \qquad \frac{N_1N_2}{ N} m v= \frac{N_1N_2}{ N} m (v_{2}-v_{1}) \label{relmom}
\\
-\Theta\qquad & \qquad  \hbar N=\hbar (N_{1} + N_{2})
\\
-\theta\qquad & \qquad \frac{\hbar n}{2}=\hbar\frac{ N_{2} - N_{1}}{2}
\,\,,
\end{align}
\label{conjecturedCanVars}
\end{subequations}
 where, following the reintroduction of dimenisonalities, we have
\begin{align}
 \theta&=\theta_{2}-\theta_{1} +\pi+\frac{m}{\hbar}\left(\frac{n}{N}b v-b V\right)  & 
 \Theta&=\frac{\theta_{1}+\theta_{2}}{2}-\frac{\pi}{2}+\frac{m}{\hbar}\left[
 -\frac{1}{4}\left(1+\frac{n^{2}}{N^{2}}\right)b v - B V + \frac{n}{2N}b V\right]\,.
\end{align}
The relative momentum in (\ref{relmom}) is defined for arbitrary $N_1$ and $N_2$; the definition $3Nmv/16$ in the main text is the particular case of $N_{1}/N_{2} = 3$ or $1/3$. %Just trying out this formulation instead of "$N_{1}=3 N_{2}$ or vice-versa".

Using Eqs.~(\ref{PoissonBracket}) and (\ref{SIvariables}), we can compute the Poisson brackets between pairs of coordinates in each row of Eqs.~(\ref{conjecturedCanVars}). We find that 
 \begin{equation}
 \left\{B,\,N m V\right\}=\left\{b,\,\frac{N_1N_2}{ N} m v\right\} =\left\{-\Theta,\,
 \hbar N\right\} =\left\{-\theta,\, \hbar n/2\right\}=  1 
 \label{PoisBrackConj}
 \end{equation}
These Poisson brackets provide the Heisenberg limits in Eqs.~\uncertaintyrels\ in the main text (see \cite{Landau}).
 
One note about the momentum uncertainties in Eqs.~\uncertaintyrels\ in the main text. The uncertainty of the relative momentum (for $N_{1}/N_{2} = 3$ or $1/3$) in Eq. (5d)  depends, in general, on the uncertainties of $N$ and $n$: $\Delta(N_1 N_2 m v/N)=3 N m  \Delta v /16+m v(5 \Delta N/4-\Delta n)$. However, our breather has $v=0$ and so the contributions of $\Delta N$ and $\Delta n$ vanish. The uncertainty of the COM momentum  in Eq. (5b) is 
$\Delta(N m V)= m N  \Delta V [1+ V \Delta N/(N \Delta V)]$. Since $\langle \Delta \hat{N}_{0}^{2}\rangle=N$ and 
$\langle \Delta \hat{V}_{0}^{2}\rangle\propto N$ (see Table~\initvariances\ in the main text), in the mean-field limit $\Delta(N m V)$ becomes independent of $N$ and $V$,  in the agreement with the Galilean invariance.

\end{document}